\documentclass[draftcls,a4paper,onecolumn]{IEEEtran}
\usepackage{segbib2,endfloat} 
\usepackage{url}
\ifCLASSINFOpdf
\usepackage[pdftex]{graphicx}
\DeclareGraphicsExtensions{.pdf,.jpg,.png}
\else
\fi
\begin{document}

\title{Deriving ice thickness, glacier volume and bedrock morphology of the Austre Lov\'enbreen (Svalbard) using Ground-penetrating Radar}

\author{\IEEEauthorblockN{A. Saintenoy\IEEEauthorrefmark{1},
J.-M. Friedt\IEEEauthorrefmark{2},
A. D. Booth\IEEEauthorrefmark{3}\IEEEauthorrefmark{6},
F. Tolle\IEEEauthorrefmark{4}, 
\'E. Bernard\IEEEauthorrefmark{4},\\
D. Laffly\IEEEauthorrefmark{5},
C. Marlin\IEEEauthorrefmark{1} and
M. Griselin\IEEEauthorrefmark{4}}\\
\IEEEauthorblockA{\IEEEauthorrefmark{1}IDES, UMR 8148 CNRS,
Universit\'e Paris Sud, 
Orsay, France\\
Email: albane.saintenoy@u-psud.fr}\\
\IEEEauthorblockA{\IEEEauthorrefmark{2}FEMTO-ST, UMR 6174 CNRS,
Universit\'e de Franche-Comt\'e,
Besan\c{c}on, France}\\
\IEEEauthorblockA{\IEEEauthorrefmark{3}Glaciology Group, Department of
  Geography, Swansea University, Swansea, Wales, UK}\\
\IEEEauthorblockA{\IEEEauthorrefmark{4}TH\'EMA, UMR 6049 CNRS,
Universit\'e de Franche-Comt\'e,
Besan\c{c}on, France}\\
\IEEEauthorblockA{\IEEEauthorrefmark{5}GEODE, UMR 5602 CNRS,
Universit\'e de Toulouse,
Toulouse, France}\\
\IEEEauthorblockA{\IEEEauthorrefmark{6}Now at: Department of Earth Science and Engineering,
  Imperial College London, South Kensington Campus, London, SW7 2AZ, UK}
}

\maketitle

\clearpage

\begin{abstract}

The Austre Lov\'enbreen is a 4.6~km$^2$ glacier on the Archipelago of
Svalbard (79$^o$N) that has been surveyed over the last 47~years in
order of monitoring in particular the glacier evolution and associated
hydrological phenomena in the context of nowadays global warming. A
three-week field survey over April 2010 allowed for the acquisition of
a dense mesh of Ground-penetrating Radar (GPR) data with an average of
14683 points per km$^2$ (67542 points total) on the glacier surface.
The profiles were acquired using a Mal\aa~ equipment with 100 MHz
antennas, towed slowly enough to record on average every 0.3~m, a
trace long enough to sound down to 189~m of ice. One profile was
repeated with 50~MHz antenna to improve electromagnetic wave
propagation depth in scattering media observed in the cirques closest
to the slopes. The GPR was coupled to a GPS system to position traces.
Each profile has been manually edited using standard GPR data
processing including migration, to pick the reflection arrival time
from the ice--bedrock interface. Snow cover was evaluated through 42
snow drilling measurements regularly spaced to cover all the glacier.
These data were acquired at the time of the GPR survey and
subsequently spatially interpolated using ordinary kriging. Using a
snow velocity of 0.22~m/ns, the snow thickness was converted to
electromagnetic wave travel-times and subtracted from the picked
travel-times to the ice--bedrock interface. The resulting travel-times
were converted to ice thickness using a velocity of 0.17~m/ns. The
velocity uncertainty is discussed from a common mid-point profile
analysis. A total of 67542 georeferenced data points with GPR-derived
ice thicknesses, in addition to a glacier boundary line derived from
satellite images taken during summer, were interpolated over the
entire glacier surface using kriging with a 10~m grid size. Some
uncertainty analysis were carried on and we calculated an averaged ice
thickness of 76~m and a maximum depth of 164~m with a relative error
of $11.9 \%$. The volume of the glacier is derived as 0.3487$\pm$0.041~km$^3$. 
Finally a 10-m grid map of the bedrock topography was
derived by subtracting the ice thicknesses from a dual-frequency
GPS-derived digital elevation model of the surface. These two datasets
are the first step for modelling thermal evolution of the glacier and
its bedrock, as well as the main hydrological network.

\bigskip

\end{abstract}

\IEEEpeerreviewmaketitle

\clearpage

\section{Introduction}

Long-term studies of the Spitsbergen Western coast glaciers reveal
that they are retreating over the last decades
\cite{hagen2003,kohler2007}. Quantification of current mass-balance
trends of these glaciers is attempted by the evaluation of surface
conditions (accumulation and ablation), basal conditions (melting or
freezing) and ice dynamics (mass movements). Surface changes can be
evaluated from digital elevation models (DEMs) derived, e.g. from photogrametric
methods applied on aerial and satellite images, surface GPS
measurements or airborne LiDAR acquisitions or ground based high
resolution photography \cite{cuffey2010,bernard2012} in addition
to {\em in situ} ablation stake network height measurements. However, the
glacier volume estimate is necessary for either ice dynamical
modelling or future mass balance scenarios.

Ground-penetrating Radar (GPR) is a geophysical tool using
radiofrequency electromagnetic waves for sounding underground
features. This method is especially efficient for mapping glaciers
thanks to the good penetration depth of the electromagnetic waves in a
low loss medium such as ice. Common-offset radar
profiling has been successfully used for evaluating ice thickness of
glaciers (e.g. \cite{hagen1991,ramirez2001,fischer2009}), deriving at a
decimetric scale the internal geometry of ice structures
\cite{hambrey2005}, locating and characterizing englacial
channels~\cite{stuart2003} and analyzing the glacier base for
determining the thermal regime \cite{murray2000,murray2009}. Multi-offset
profiles are acquired for getting a wave velocity estimate or the
water content variations of the glacier ice~\cite{murray2007}. It is
striking to see the evolution in the radar surveys since the 1990s when
measurement positioning was achieved using compass and visual
navigation on the glacier and the main source of
error in the ice thickness estimation was considered to be $\pm$ 10 m
mostly due to the digitizing of the profiles~\cite{hagen1991}. High resolution,
real time positioning capability as provided by GNSS and, in our case, GPS, provides 
the mandatory tool for high resolution bedrock mapping on challenging terrain.

The Austre Lov\'enbreen is a northward-flowing valley glacier situated
in the Br\o gger peninsula, Spitsbergen, Norway (79$^o$N)
\cite{mingxing2010,bernard2011}. It extends from an altitude of 100~m
to 560~m above sea level. The mean annual precipitation is 391~mm and
its mean annual temperature from 1969 to 1998 is -5.77$^o$C (source
DNMI at http://eklima.met.no). Thanks to the geological configuration
of its basin, all runoff water is concentrated into two channels. With
this specific hydrological configuration and being near the former
mining town of Ny-\AA lesund, this site has been the focus of intense
scrutiny since the 1960s. A summary of historical dataset since 1962,
used for elevation models on this glacier as well as their relevance
to evaluate mass balance is described in \cite{friedt2012}. The
neighboring glacier, Midtre Lov\'enbreen, has been extensively studied
as well. It is known to be polythermal on the basis of radio echo
sounding~\cite{hagen1991,bjornsson1996,rippin2003}. A detailed
description of its structure and dynamics can be found in
\cite{hambrey2005}. Additionally to a high frequency GPR survey, a
seismic reflection survey allowed for determining the properties of
the bed material~\cite{king2008}.

In this paper we present results of a high density coverage GPR survey
(120 profiles resulting in 67542 ice thickness measurements) of the
Austre Lov\'enbreen. We first show some internal structures observed
on selected radargrams, then present the ice volume estimation and
finally the glacier substratum topography. We discuss the different
sources of uncertainties in those two data sets.

\section{Data collection and processing}

\begin{figure}[h!tb]
\centering \includegraphics[width=15cm]{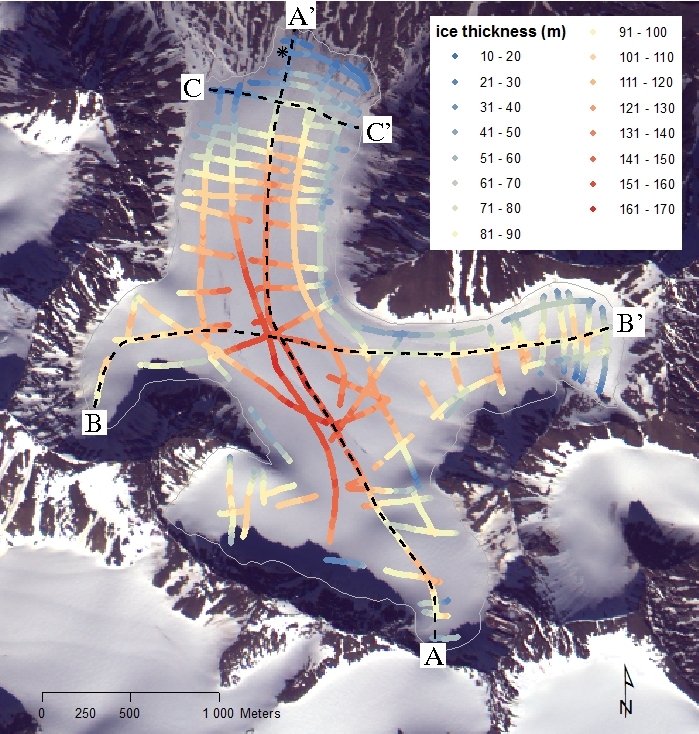}
\caption{GPR profiles over the Austre Lov\'enbreen (background image
  copyright FORMOSAT). The color scale indicates the ice thickness
  measured on each profiles. Dashed lines indicate GPR transects
  displayed on Fig.~\ref{longitu-mid} to~\ref{prof-tongue}. The
  symbol $\ast$ indicates the CMP position.}
\label{prof-location}
\end{figure}

We used a Mal\aa~Ramac GPR operating at 50 and 100 MHz to collect more than
70~km of mono-offset profiles (Fig.~\ref{prof-location}) over the
surface of the Austre Lov\'enbreen (Svalbard) during 3 weeks in April
2010. Both the 50 MHz and 100 MHz antenna data, corresponding to a
nominal wavelength in ice of 3.4~m and 1.7~m respectively, were
collected in the form of 2806 samples within a time window 2.224
$\mu$s. All data were stacked 8 times on collection. Positioning of all GPR
mono-offset profiles was done using a Globalsat ET-312
Coarse/Acquisition (C/A) code GPS receiver connected to the
control unit of the GPR, set to 1 measurement per second while two
operators were pulling the device at a comfortable walking pace. A
trace was acquired every 0.5~s, and the average distance between
traces was later calculated at 0.3~m.

\begin{figure}[h!tb]
\centering \includegraphics[width=15cm]{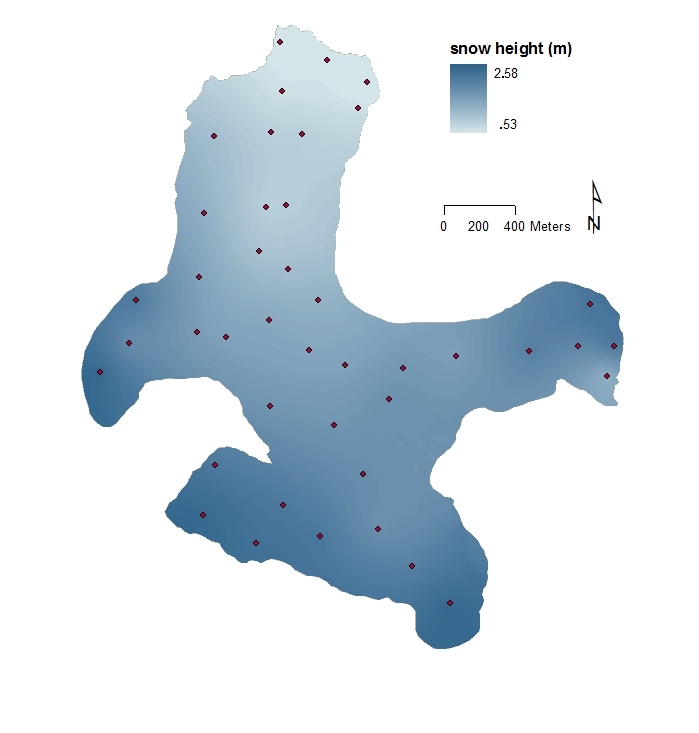}
\caption{Snow thickness map interpolated from 42 snow drilling
  measurements indicated by the red dots. The mean snow thickness was
  evaluated to 1.67 m.}
\label{snow-map}
\end{figure}

Snow cover was evaluated through 42 snow drilling measurements
regularly spaced to cover all the glacier. These data were acquired at
the time of the GPR and dual GPS measurements and subsequently
interpolated using ordinary kriging. The resulting snow thickness map
is shown on Fig.~\ref{snow-map}. The measurement root mean square
error is 20~cm~\cite{webster2001}. The average snow thickness over the
entire glacier on April 2010 was estimated to 1.67~m.

\begin{figure}[h!tb]
\centering
\includegraphics[width=15cm]{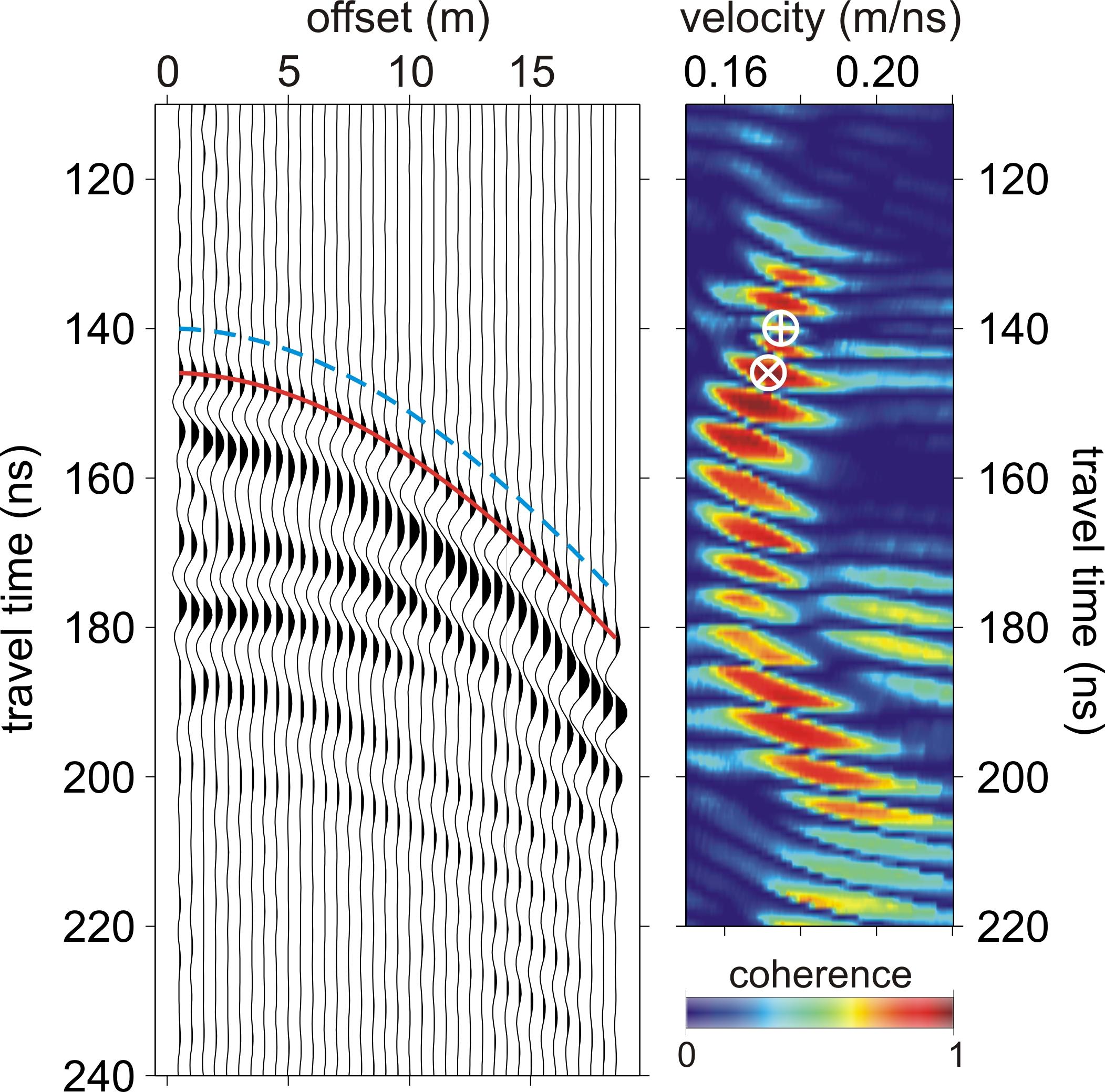}
\caption{Coherence analysis of CMP data  located on the glacier snout
  (Fig. ~\ref{prof-location}). Left: CMP data and reflection
  trajectories, as defined from coherence analysis (right). The lower
  hyperbola (red) corresponds to the lower coherence pick, and is
  interpreted as the reflection from the ice--bedrock interface. After
  application of backshifting (Booth et al., 2010), we define the
  upper hyperbola (blue, dashed) and coherence pick as a better
  approximation to first-break travel-times.}
\label{CMP}
\end{figure}

In addition to the mono-offset profiles, one Common Mid-Point (CMP)
gather was acquired on the glacier snout using the 100 MHz antennas
(Fig~\ref{CMP}). The initial separation between antennas was 0.5~m,
with a spatial stepsize of 0.5~m. CMP data were interpreted using
coherence analysis, defined equivalently to semblance but using an
analysis window of one temporal sample (here, 0.8~ns). The basal
reflection exhibits a velocity of 0.1715~m/ns (red trajectory in CMP
gather, lower pick in coherence panel), but coherence delivers a
root-mean-square velocity that is biased systematically slow with
respect to its true value~\cite{booth2010}. This occurs because true
velocity is only expressed by wavelet first-breaks, yet these are zero
amplitude hence cannot produce a coherence response. We therefore use
the coherence response to simulate first-break travel-times, using the
'backshifting' method of Booth et al.~\shortcite{booth2010}, and
obtain an RMS velocity of 0.1747~m/ns and a travel-time to the base of
the ice of 140.0~ns (blue trajectory in CMP gather, upper pick in
coherence panel).

This RMS velocity is then converted to interval velocity using Dix's
equation~\cite{dix55}. At the location of the CMP acquisition, the
glacier was covered by 0.7~m of snow, which we assume to have a
velocity of 0.22~m/ns \cite{murray2007} and, hence, the two-way travel-time to
the base of the snow is 6.3~ns. Substituting our velocity-time model
into Dix's Equation gives 0.1723$\pm$0.0021~m/ns as the interval
velocity through the ice, and a local ice thickness of 10.21$\pm$0.16~m. 
The uncertainty in these values is obtained by considering the
resolution of coherence analysis~\cite{booth2011}, and is therefore
representative of the error between a given coherence pick and its
true velocity value.

Successive depth conversions are made with a velocity value of
0.17~m/ns, which represents the lower-bound of the error in interval
velocity. We choose this value since the volumetric content of air is
likely to decrease in the thicker parts of the
glacier~\cite{gusmeroli2010} hence we anticipate that a slower
velocity is more widely representative. Although CMP surveys over the
thickest ice could confirm this, the fiber optic cables of our GPR
system were only 20~m long, reducing our maximum offset-to-depth ratio
and thereby producing a poor coherence response. Finally, we
will use the velocity derived from the 100 MHz dataset
to depth-convert 50~MHz records. Ice is weakly dispersive:
across the range 1-100~MHz, relative dielectric permittivity decreases
by 0.04~\cite{Dowdswell2004}. Accordingly, in terms of
propagation velocity, if our 100~MHz wavelet travels at 0.1700~m/ns, a
50~MHz wavelet travels at 0.1695~m/ns, a difference that we consider
negligible in depth conversion.

Mono-offset GPR data have been processed using Seismic Unix software
\cite{cohen2010,stockwell1999}. A residual median filter was applied
in vertical direction using a time window corresponding to the cut-off
frequency of 50 MHz, each trace has been normalized to its root mean
square value and bandpass filtered. Each profile was chopped above the
arrival time of the minimum amplitude of the direct air wave (manually
selected). Based on the ET312 C/A GPS information, the mean
  distance $a$ between traces is computed. Equidistant trace
  positioning is achieved by searching for the acquired trace located
  closest to a periodic grid of period $a$. The obtained profiles
  have then been migrated using a Stolt algorithm with a velocity of
0.17 m/ns. When needed for visualization, elevation correction was
implemented using the altitude given by the ET312 C/A GPS.

During the GPR survey, a dense elevation map was performed using GPS
measurements with a snowmobile: a Trimble Geo-XH dual frequency
receiver, with electromagnetic delay correction post-processing using
the nearby ($<$10~km away) Ny-\AA lesund reference dataset, provided
the raw data to generate a DEM of the glacier after interpolation of
the dataset. Data processing is performed in two steps. First the
  ice thickness is derived from GPR profiles, with removal of the snow
  thickness contribution. In a second step, the bedrock surface is
  interpolated and located in space by subtracting the ice thickness
  from the surface DEM. 

\section{Glacier structures}

\begin{figure}[h!tb]
\centering
\includegraphics[angle=90,height=23cm]{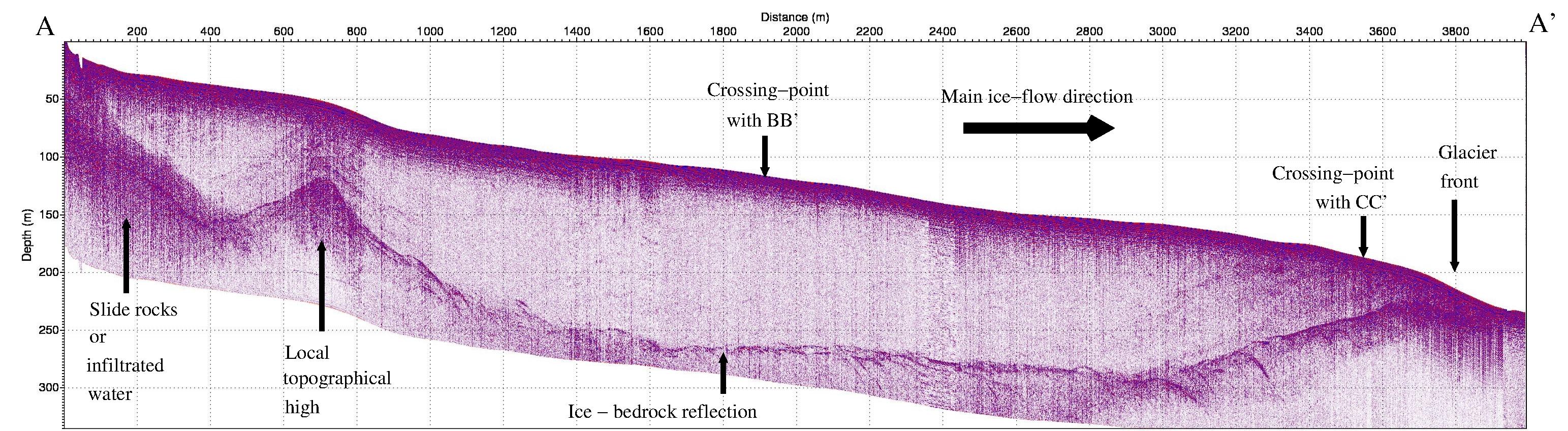}
\caption{Radargram AA' acquired along the glacier axis with 100 MHz
  antennas including topography corrections.}
\label{longitu-mid}
\end{figure}

\begin{figure}[h!tb]
\centering
\includegraphics[angle=90,height=23cm]{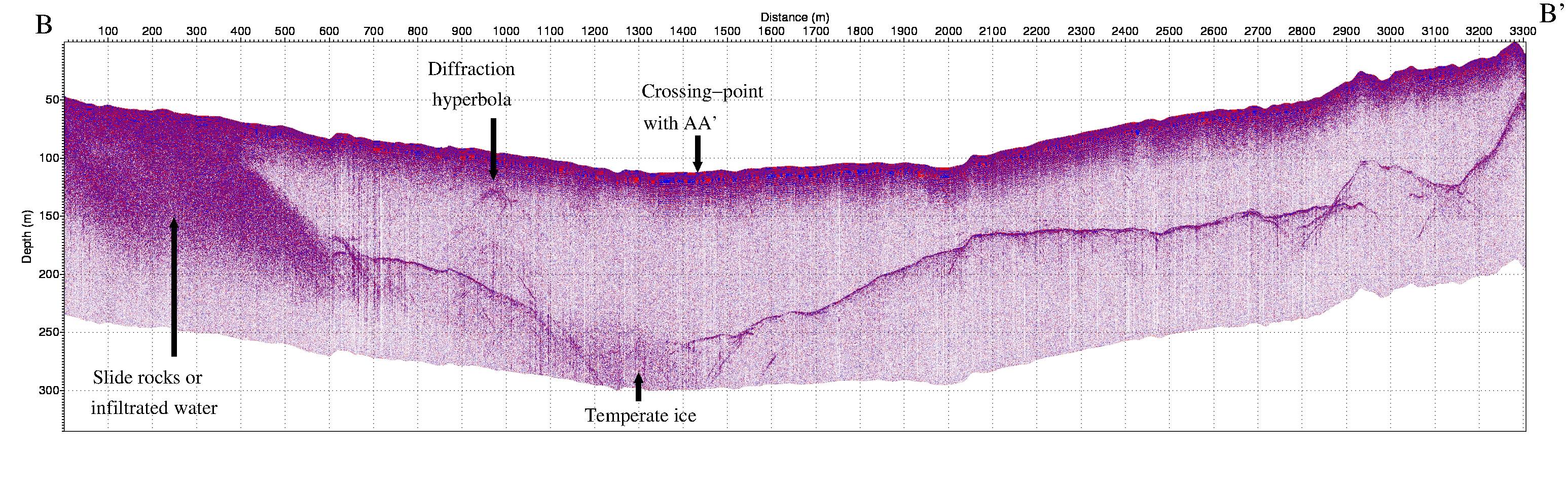}
\caption{Radargram BB' acquired across the glacier axis with 100 MHz
  antennas (non migrated).}
\label{prof-accross}
\end{figure}

\begin{figure}[h!tb]
\centering
\includegraphics[width=15cm]{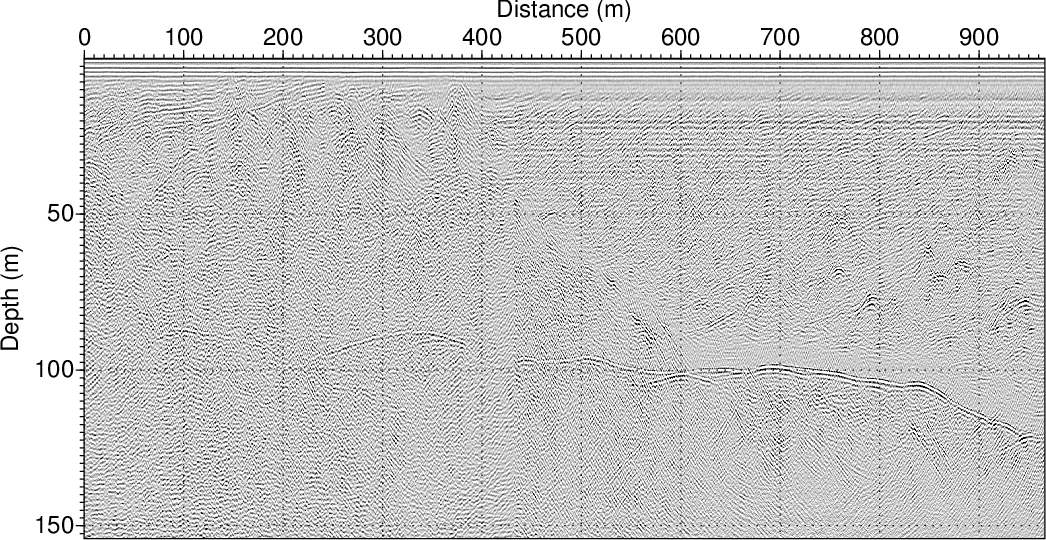}
\caption{Repetition of the first 900~m of profile BB' with 50 MHz
  antennas (after Stolt migration using a velocity of 0.17 m/ns, with AGC gain but no
  topographic corrections).}
\label{50MHz}
\end{figure}

\begin{figure}[h!tb]
\centering
\includegraphics[width=15cm]{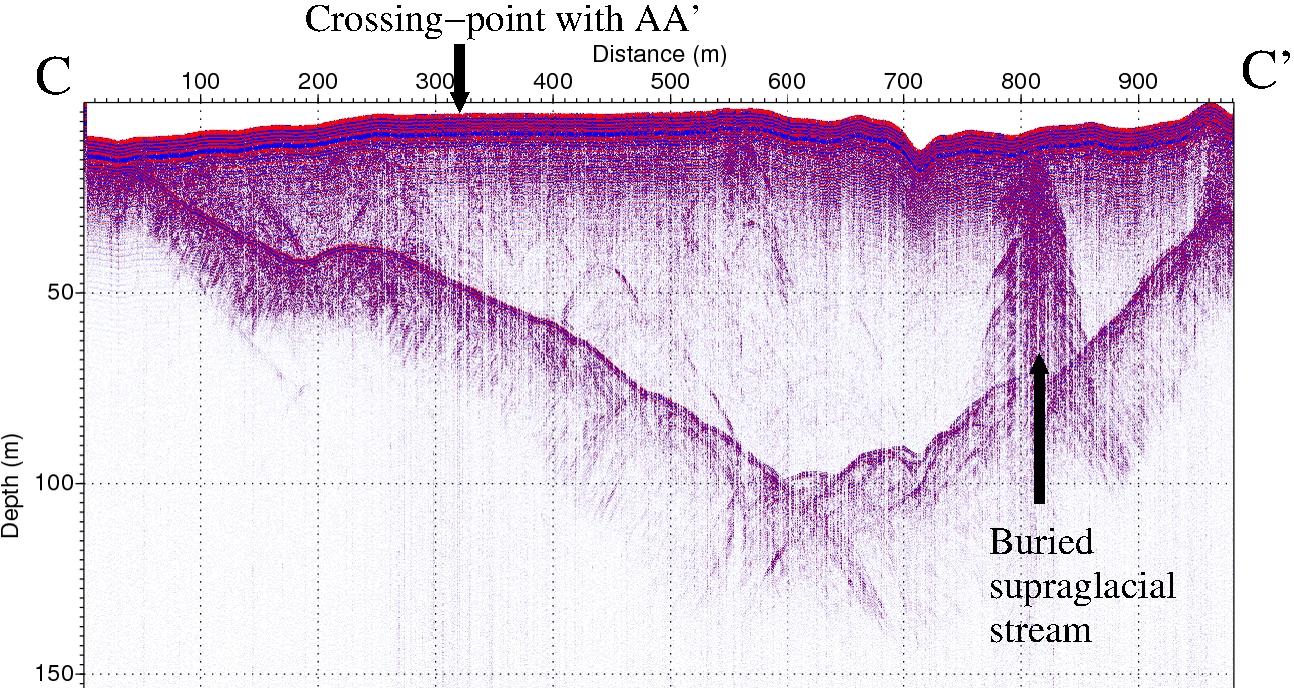}
\caption{Radargram CC' acquired across the glacier tongue with 100 MHz
  antennas (non migrated).}
\label{prof-tongue}
\end{figure}

For giving an insight on our GPR data quality, four processed
radargrams are shown on Fig.~\ref{longitu-mid}, \ref{prof-accross}, \ref{50MHz}
and~\ref{prof-tongue}. AA' was acquired along the glacier central axis
toward North while BB' was acquired from West to East across the
glacier (see Fig.~\ref{prof-location}). CC' was acquired across
  the glacier tongue.

Along AA', the strong continuous reflection is interpreted as the
ice--bedrock interface. {The main ice-flow direction is South to
  North~\cite{mingxing2010}.} At the beginning of the profile,
multiple scattering occurs partially masking the ice--bedrock
interface, preventing sometime the picking of the arrival time of the
radar reflection on this interface. Similar zones are observed only on
the western upper side of the glacier as in profile BB' beginning. We
interpret these as reflections on fallen rocks incorporated into the
ice, or infiltrated water. Around 700~m, the bedrock topography rises
by 50~m over a distance of 200~m creating a local topographical high.
This feature may be related to the geology of the area: a thrust fault
between the Welderyggen thrust sheet and the Nielsenfjellet thrust
sheet is indicated in the southern part of the Lov\'enbreen glacier in
geological maps~\cite{hjelle1993,saalmann2002}. Heading farther north
the bedrock surface is easy to follow all the way down to the glacier
frontal moraine. At 3000~m along this profile, the ice thickness
decreases as the bedrock topography rises 70~m. The same trends are
observed in other parallel profiles.

On BB' radargram, the bedrock reflection is very clear except in two
areas. In the middle part of the glacier (around 1300 m), an area with
increased scattering appears in the deepest part of the glacier, and
we attribute this to the presence of temperate ice as described in the
neighboring glacier~\cite{hagen1991,moore99,king2008}. {This multiple
scattering area prevented us from picking the ice--bedrock interface
reflection resulting into a gap in ice thickness estimates as visible
on Fig.~\ref{prof-location}. In this figure, other gaps in the center
part of the glacier result from the same difficulty to pick the
ice--bedrock interface due to high scattering zones, giving thus an
idea of the horizontal extension of the probable temperate ice.}

On the first 500~m of BB', many scatterers are again observed,
associated with either fallen rocks or infiltrated water given the
proximity to the surrounding mountain side. {The bedrock reflection
disappears among all the scatterers but it becomes detectable in the
profile acquired in this area using 50~MHz antenna (Fig.~\ref{50MHz}).
This 50~MHz migrated profile was used to pick the ice--bedrock
interface instead of the first 900~m of profile BB'}. At 1000~m along
the profile, 30~m deep, some large hyperbolae are attributed to buried
englacial channels. {Our data set does not present parallel profiles
close enough to BB' to determine the horizontal extension of this
channel.}

Fig.~\ref{prof-tongue} shows one processed profile across the
glacier tongue along the profile CC' of Fig.~\ref{prof-location}.
This profile crosses a buried supraglacial stream, evident on the
satellite image of June 26th 2007, copyright FORMOSAT.
Where this stream intersects CC', at around 800 m,
the radargram shows many diffraction hyperbolae. This feature can be
observed an all radargrams that cross the stream.

\section{Ice volume estimation}

The boundary of the glacier (grey line in Fig.~\ref{prof-location}),
14143~m long, was drawn on a summer 2009 FORMOSAT image. Whenever
visible, rimaye (bergschrunds) were considered as the limit between
the glacier and slopes. Moreover, slope angles were derived and used
to differentiate steep angle slopes and low angle glacier. Field
knowledge and direct local GPS measurements were of great help as
well. Visual inspection of all these elements allowed us to determine
as precisely as possible the limits of the glacier. We estimate that
the glacier boundary is identified with a $\pm$10~m uncertainty. The
area of the glacier is thus measured to be 4.6$\pm$0.28~km$^2$. We
have decided to define a null ice thickness on the boundary. We will
see that this choice will not significantly affect the ice volume
estimate: assuming a maximum of 20~m ice thickness error along this
boundary, the volume contribution is 0.0056~km$^3$ (1.6\% relative
error).

In every migrated GPR profile, the arrival time of the basal
reflection was picked using Reflexw software~\cite{sandmeier2007}. No
picking was done where the ice--bedrock interface was not clear. The
two-way travel time is translated into ice thickness using 0.17~m/ns
velocity as derived earlier from CMP analysis: the uncertainty on this
velocity contributes to 1.2 \% uncertainty on the glacier volume (Fig.
\ref{interp_ice_thickness}). The snow layer contribution (Fig.
\ref{snow-map}) to the radar wave propagation duration is removed by
subtracting its corresponding time delay assuming a velocity of 0.22
m/ns. The uncertainty on the snow thickness of 20~cm contributes to a
volume of 9.2$\times 10^{-4}$~km$^3$ (0.3\% relative error).

The analysis ended-up with a total of 67542 georeferenced data points
with GPR-derived ice thickness. All ice-thickness measurements were
interpolated over the entire glacier surface using a kriging method
onto a 10~m grid (Fig.~\ref{interp_ice_thickness}). The ice volume is
0.3487~km$^3$. Notice that working on non migrated data as
in Saintenoy et al.~\shortcite{saintenoy2011b} yields a volume of 0.3427~km$^3$, or a 1.1~\%
error with respect to the volume derived from migrated data.

\begin{figure}[h!tb]
  \centering
\includegraphics[width=12cm]{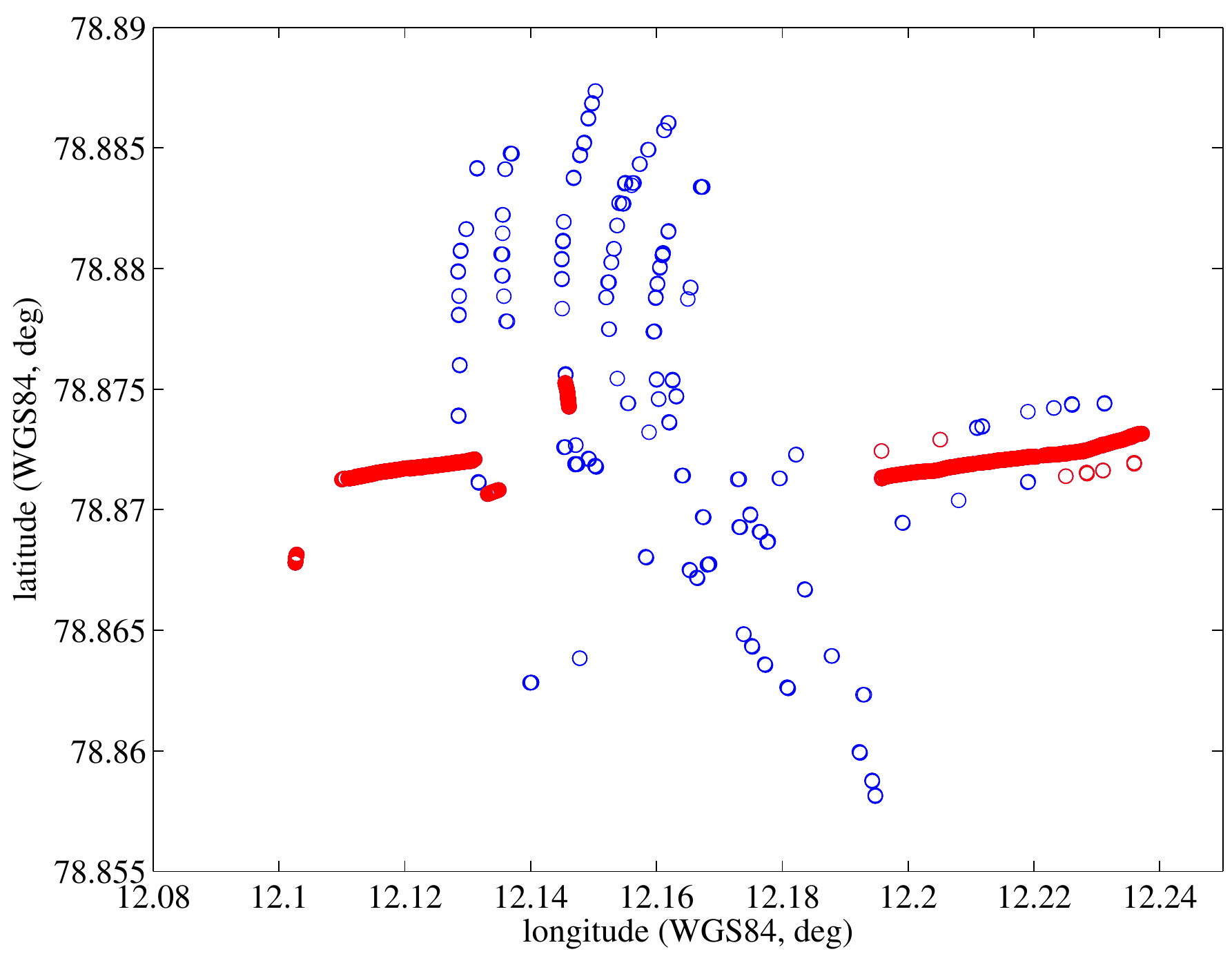} \\
\includegraphics[width=12cm]{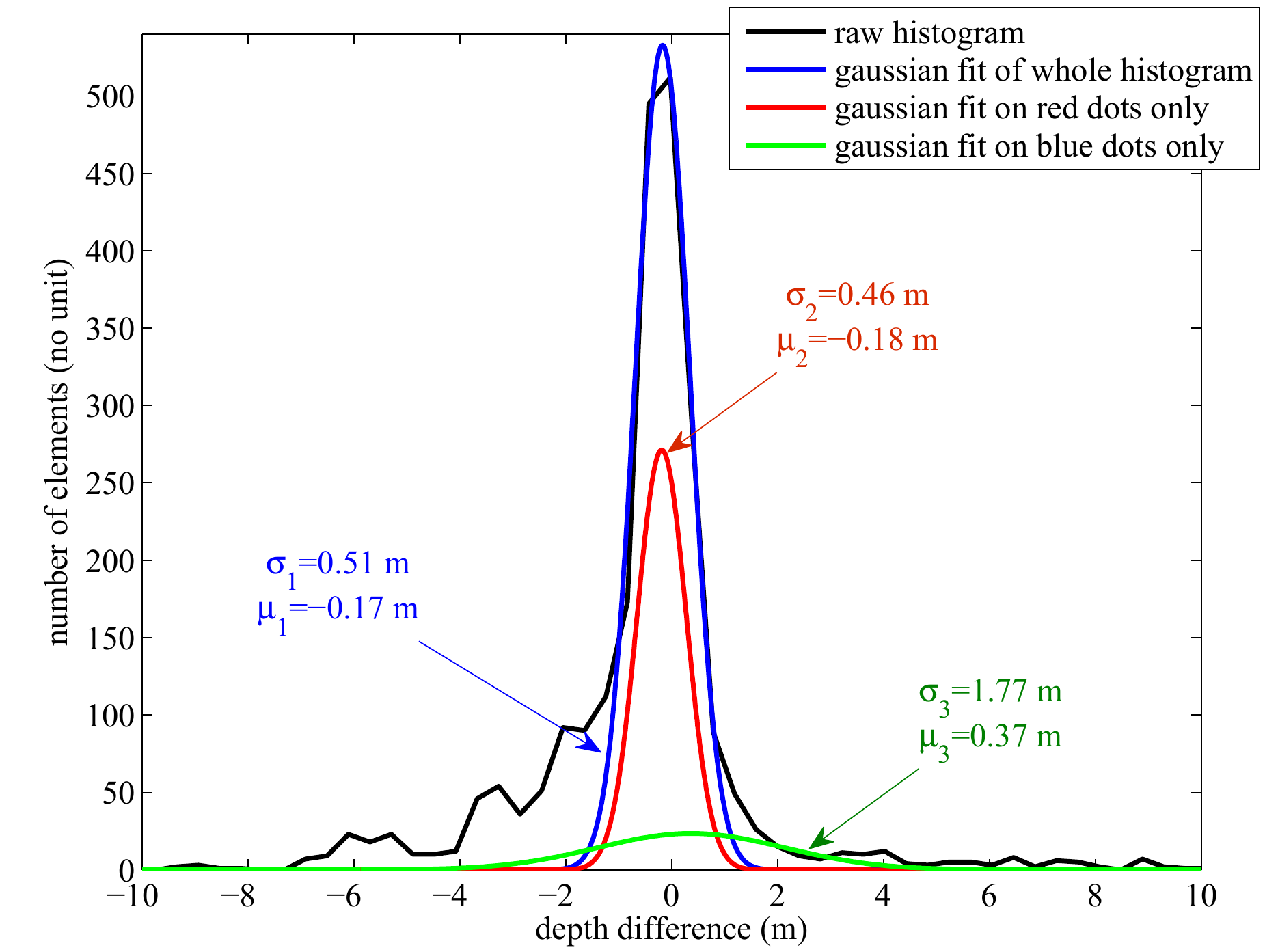} 
\caption{Top: map of the analyzed intersection points with \em
    closest points located less than 
3~m from each other at each GPR track intersection. Blue is all intersection points, red
is a particular case of two measurements performed several days apart but following the
exact same path over the glacier (tracks left in the snow).
Bottom: histogram of the depth difference distribution. The analysis was performed on
various subsets of the intersection dataset, with the standard
deviation $\sigma$ and the mean value $\mu$ of the 
gaussian fit indicated for each contribution.}
\label{histo}
\end{figure}

Depth estimate quality assessment was performed by analyzing the error
between ice-thickness estimates from closely-separated traces in
distinct profiles: the thickness difference between the closest points lying less
than 3~m apart was computed and the histogram of the ice-thickness
distribution is plotted (Fig.~\ref{histo}). A gaussian fit of each
histogram is performed using a constrained nonlinear optimization
method: we analyze the whole dataset including all traces
intersections (blue dots) and separately the particular case of five
transects acquired 3 to 5 days apart but following the same path (snow
tracks).
The global histogram exhibits a standard deviation of 0.51~m and a
negligible mean value of -0.17~m. These results, suggesting a better
agreement than other analysis found in the
literature~\cite{fischer2009,hagen1991}, is however optimistic by
including the five repeated transects with standard deviation 0.46~m
and mean value of -0.18~m. Using only intersections of traces crossing
at high angles by excluding the five repeated transects (Fig.
\ref{histo}, top, red points), the standard deviation of the histogram
increases to 1.77~m with a mean value of 0.37~m.

This histogram of the ice thickness differences at intersections
analysis is consistent with the result of the surface interpolation by
kriging, which provides an estimate of the root-mean square error
between experimental data and the interpolated surface of 0.7~m.
However the interpolation of ice thickness outside of the tracks
yields the largest source of uncertainty, as provided by the kriging
prediction standard error map shown on Fig.~\ref{ice_thickness_error},
with a 11.5\% contribution to the ice volume calculation
(corresponding to an average error of 8.72~m on the interpolated ice
thickness).

As a result, the ice volume was estimated to 0.3487 $\pm$
0.041~km$^3$, with all contributions to the uncertainty summarized in
Table~\ref{table1}. This result is to be compared with the empirical
formula found in Hagen et al.~\shortcite{atlas1993} for outlet
glaciers whose area $A$ exceeds 1~km$^2$: the mean depth is estimated
as $D=33\log(A)+25$. In our case, $A=4.6\pm 0.28$~km$^2$ yields a mean
depth of 75$\pm$2~m, surprisingly close to the 76~m we found from our
analysis.

\begin{table}[h!tb]
\begin{center}
  \begin{tabular}{|c|c|c|} \hline
Cause of error & Volume & Relative error \\ \hline \hline
    Ice thickness: $\pm 1.77$~m & 0.008~km$^3$& 2.3 \%\\
    Glacier area & 0.0056~km$^3$ & 1.6 \%\\
    Snow thickness & 9.2 $\times 10^{-4}$~km$^3$ & 0.3 \% \\
    Electromagnetic wave velocity & 0.0042~km$^3$ & 1.2 \% \\ \hline 
Interpolation error & 0.040~km$^3$ & 11.5 \% \\ \hline
  \end{tabular}
\end{center}
  \caption{Summary of contributions to glacier volume estimation
    error. The sum of all errors yields to a 11.9~\% accuracy.}
  \label{table1}
\end{table}

\section{Bedrock digital elevation model}

\begin{figure}[h!tb]
\centering
\includegraphics[width=15cm]{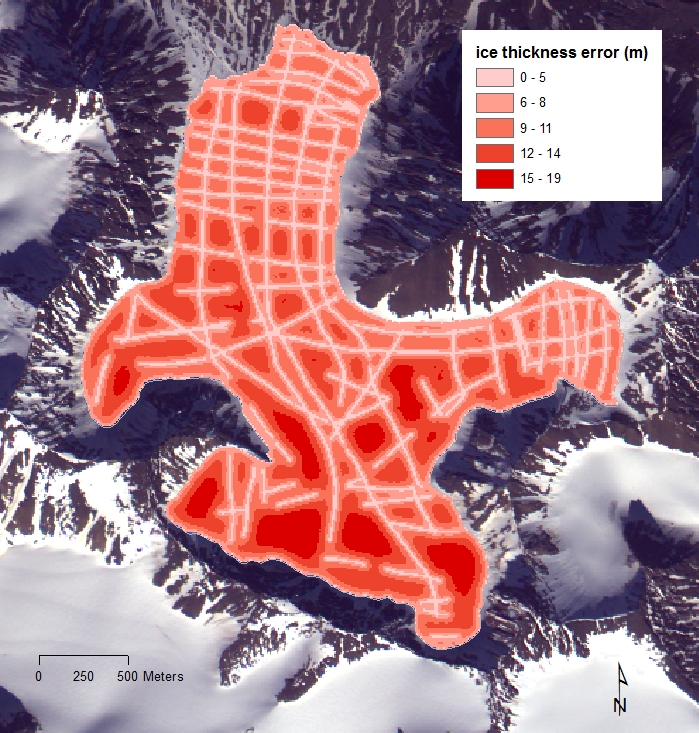}
\caption{Prediction standard error map of interpolated ice thicknesses (background image copyright FORMOSAT).}
\label{ice_thickness_error}
\end{figure}

The Coarse/Acquisition (C/A) code GPS receiver that was used when GPR
data was acquired is in the range of a 3~m standard deviation in
latitude and longitude but displays an unacceptable vertical accuracy
with respect to the DEM resolution. Therefore, only dual-frequency
acquired GPS altitude measurements were used as DEM reference for
bedrock positioning. The accuracy of surface DEM is discussed in
detail in Friedt et al. (2012). The same uncertainty analysis carried
on the dual-frequency GPS measurements yields an altitude distribution
with a standard deviation less than 0.6~m. Uncertainties on ice and
snow thicknesses as well as on the electromagnetic wave velocity
estimation, sum up to 2.6~\% error on the 76~m-mean depth glacier
thickness, or 2~m. Thus, considering a 0.6~m standard deviation on the
DEM height, the bedrock
topography uncertainty (standard deviation of altitude error) is 2.6~m over
the measurement points. The error analysis from the kriging interpolation
rises this error on the interpolated areas to 19.6~m for areas far from any
experimental dataset (cf Fig.~\ref{ice_thickness_error}).

\begin{figure}[h!tb]
\centering
\includegraphics[width=15cm]{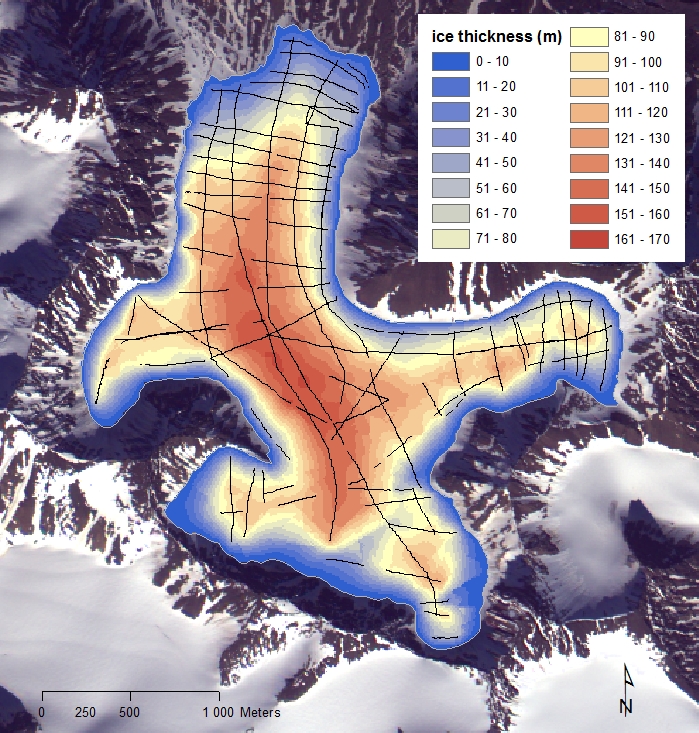}
\caption{Interpolated ice thickness with GPR transects in black lines (background image copyright FORMOSAT).}
\label{interp_ice_thickness}
\end{figure}

\begin{figure}[h!tb]
\centering
\includegraphics[width=15cm]{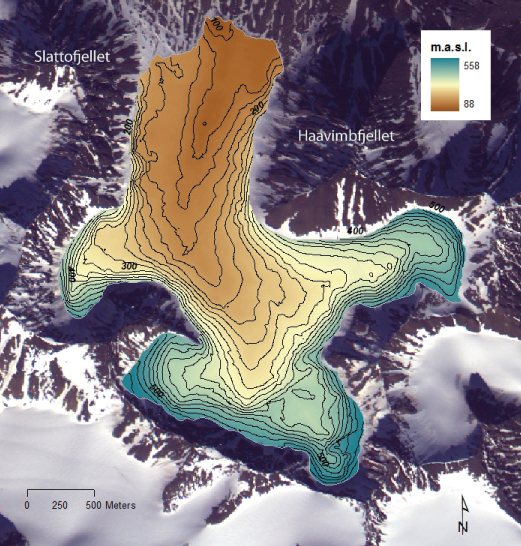}
\caption{DEM of the glacier substratum with 20-m spaced contour lines (background image copyright FORMOSAT).}
\label{substra}
\end{figure}

Figs.~\ref{interp_ice_thickness} and~\ref{substra} show the asymmetry
of the bedrock underneath the ice on the glacier snout. The substratum
is deeper on the easter side of the glacier. Furthermore, the
ice--bedrock appears convex (bulging outward) on the western side and
concave (hollowed inward) on the eastern side as seen on the GPR
profiles acquired across the glacier snout (Fig.~\ref{prof-tongue}).
This observation is consistent with a difference in the hardness of
the underlying rock, and possibly to the transform fault presented in
the geological map of Saalmann and Thiedig~\shortcite{saalmann2002} in
between the Slatto and the Haavimb summits (Fig.~\ref{substra}).

\section{Conclusion}

A high resolution mapping by 100-MHz and 50-MHz GPR of a polar glacier
provides a detailed bedrock topography information. While the average
ice thickness of 76~m is consistent with empirical data derived from
glaciers in the Svalbard area, the high resolution dataset obtained by
walking yields a rich information including subsurface structures
(crevasse fields, bedieres, supraglacial stream) and ice volume
distribution amongst the various glacier substructures (cirques). The
resulting volume is estimated to 0.3487 $\pm$ 0.041~km$^3$, with the
main source of error being the interpolation uncertainty of the ice
thickness between tracks.

Such volume distribution provide the basic input for further mass balance 
investigations.  
Furthermore, high density GPR data coverage coupled to accurate DEM obtained
by dual frequency GPS provides a map of the bedrock following an interpolation
by kriging. This bedrock digital elevation model exhibits asymmetric features
consistent with geological structures (faults) in the area.
Bedrock morphology can now be used to investigate subglacial water
flow paths, to be improved by considering the influence of ice pressure.

\section*{Acknowledgment}

This program was funded by the ANR program blanc-0310, the IPEV program
304 and the CNRS-GDR 3062 Mutations polaires. Adam Booth is supported
by the Leverhulme-funded GLIMPSE project. The authors would like to
thank AWIPEV for the logistical support in Ny-\AA lesund, Tavi Murray
for her much helpful comments to realize this work, Nerouz Boubaki and
Emmanuel L\'eger for picking some GPR data and M\'elanie Quenet for
pointing out some references on the geology of the area.


\begin{thebibliography}{}

\end{thebibliography}


\clearpage

\begin{thebibliography}{30}

\bibitem[\protect\citeauthoryear{Bernard \bgroup et al.\egroup }{in
  press}]{bernard2012}
Bernard, E., Friedt, J.-M., Tolle, F., Griselin, M., Martin, G., Laffly, D.,
  and Marlin, C., in press, Monitoring seasonal snow dynamics using ground
  based high resolution photography ({A}ustre {L}ov\'enbreen, {S}valbard,
  79$^o$ {N}): ISPRS Journal of Photogrammetry and Remote Sensing.

\bibitem[\protect\citeauthoryear{Bernard}{2011}]{bernard2011}
Bernard, E., 2011, Les dynamiques spatio-temporelles d'un petit hydrosyst\`eme
  arctique~: approche nivo-glaciologique dans un contexte de changement
  climatique contemporain (bassin du glacier {A}ustre {L}ov\'enbreen,
  {S}pitsberg, 79 $^o$ {N}): Ph.D. thesis, Universit\'e de Franche-Comt\'e,
  Besan\c con.

\bibitem[\protect\citeauthoryear{{Bj{\"o}rnsson} \bgroup et al.\egroup
  }{1996}]{bjornsson1996}
{Bj{\"o}rnsson}, H., {Gjessing}, Y., {Hamran}, S., {Ove Hagen}, J.,
  {Liest{\o}l}, O., {P{\'a}lsson}, F.,  and {Erlingsson}, B., 1996, {The
  thermal regime of sub-polar glaciers mapped by multi-frequency radio-echo
  sounding}: Journal of Glaciology, {\bf 42}, 23--32.

\bibitem[\protect\citeauthoryear{Booth \bgroup et al.\egroup
  }{2010}]{booth2010}
Booth, A.~D., Clark, R.,  and Murray, T., 2010, Semblance response to a
  ground-penetrating radar wavelet and resulting errors in velocity analysis:
  Near Surface Geophysics, {\bf 8}, no. 3, 235--246.

\bibitem[\protect\citeauthoryear{Booth \bgroup et al.\egroup
  }{2011}]{booth2011}
Booth, A.~D., Clark, R.,  and Murray, T., 2011, Influences on the resolution of
  {GPR} velocity analyses and a {M}onte {C}arlo simulation for establishing
  velocity precision: Near Surface Geophysics, {\bf 9}, no. 5, 399--411.

\bibitem[\protect\citeauthoryear{Cohen and Stockwell}{2011}]{cohen2010}
Cohen, J., and Stockwell, J.
\newblock {CWP/SU}: {S}eismic {U}n*x {R}elease {N}o. 42: an open source
  software package for seismic research and processing:.
\newblock www.cwp.mines.edu/cwpcodes, 2011.

\bibitem[\protect\citeauthoryear{Cuffey and Paterson}{2010}]{cuffey2010}
Cuffey, K.~M., and Paterson, W. S.~B., 2010, The {P}hysics of {G}laciers:
  Boston, Elsevier, fourth edition.

\bibitem[\protect\citeauthoryear{Dix}{1955}]{dix55}
Dix, C.~H., 1955, Seismic velocities from surface measurements: Geophysics.

\bibitem[\protect\citeauthoryear{Dowdswell and Evans}{2004}]{Dowdswell2004}
Dowdswell, J.~A., and Evans, S., 2004, Investigations of the form and flow of
  ice sheets and glaciers using radio-echo sounding: Reports on Progreess in
  Physics,  1821--1861.

\bibitem[\protect\citeauthoryear{{Fischer}}{2009}]{fischer2009}
{Fischer}, A., 2009, {Calculation of glacier volume from sparse ice-thickness
  data, applied to {S}chaufelferner, {A}ustria}: Journal of Glaciology, {\bf
  55}, no. 191, 453--460.

\bibitem[\protect\citeauthoryear{Friedt \bgroup et al.\egroup
  }{2012}]{friedt2012}
Friedt, J.-M., Tolle, F., Bernard, E., Griselin, M., Laffly, D.,  and Marlin,
  C., 2012, Assessing the relevance of digital elevation models to evaluate
  glacier mass balance: application to {A}ustre {L}ov\'enbreen ({S}pitsbergen,
  79$^o${N}): Polar Record, {\bf 48}, no. 244, 2--10.

\bibitem[\protect\citeauthoryear{Gusmeroli \bgroup et al.\egroup
  }{2010}]{gusmeroli2010}
Gusmeroli, A., Murray, T., Jansson, P., Pettersson, R., Aschwanden, A.,  and
  Booth, A.~D., 2010, Vertical distribution of water within the polythermal
  {S}torglaciaren, {S}weden: Journal of Geophysical Research, {\bf 115},
  F04002.

\bibitem[\protect\citeauthoryear{Hagen and S{\ae}trang}{1991}]{hagen1991}
Hagen, J.~O., and S{\ae}trang, A., 1991, Radio-echo soundings of sub-polar
  glaciers with low-frequency radar: Polar Research, {\bf 9}, no. 1, 99--107.

\bibitem[\protect\citeauthoryear{Hagen \bgroup et al.\egroup
  }{1993}]{atlas1993}
Hagen, J.~O., Liest{\o}l, O., Roland, E.,  and J{\o}rgensen, T., 1993, Glacier
  {A}tlas of {S}valbard and {J}an {M}ayen: Norsk Polarinstitutt.

\bibitem[\protect\citeauthoryear{Hagen \bgroup et al.\egroup
  }{2003}]{hagen2003}
Hagen, J.~O., Kohler, J., Melvold, K.,  and Winther, J.-G., 2003, Glaciers in
  {S}valbard: mass balance, runoff and freshwater flux: Polar Research, {\bf
  22}, 145--159.

\bibitem[\protect\citeauthoryear{{Hambrey} \bgroup et al.\egroup
  }{2005}]{hambrey2005}
{Hambrey}, M.~J., {Murray}, T., {Glasser}, N.~F., {Hubbard}, A., {Hubbard}, B.,
  {Stuart}, G., {Hansen}, S.,  and {Kohler}, J., 2005, {Structure and changing
  dynamics of a polythermal valley glacier on a centennial timescale: Midre
  Lov{\'e}nbreen, Svalbard}: Journal of Geophysical Research (Earth Surface),
  {\bf 110}, 1006--+.

\bibitem[\protect\citeauthoryear{Hjelle}{1993}]{hjelle1993}
Hjelle, A., 1993, Geology of {S}valbard:, volume~7 Oslo: Norsk Polar Institute.

\bibitem[\protect\citeauthoryear{{King} \bgroup et al.\egroup
  }{2008}]{king2008}
{King}, E.~C., {Smith}, A.~M., {Murray}, T.,  and {Stuart}, G.~W., 2008,
  {Glacier-bed characteristics of {M}idtre Lov{\'e}nbreen, {S}valbard, from
  high-resolution seismic and radar surveying}: Journal of Glaciology, {\bf
  54}, 145--156.

\bibitem[\protect\citeauthoryear{Kohler \bgroup et al.\egroup
  }{2007}]{kohler2007}
Kohler, J., James, T.~D., Murray, T., Nuth, C., Brandt, O., Barrand, N.~E.,
  Aas, H.~F.,  and Luckman, A., 2007, Acceleration in thinning rate on
  {W}estern {S}valbard glaciers: Geophysical Research Letters, {\bf 34}, 1--5.

\bibitem[\protect\citeauthoryear{Mingxing \bgroup et al.\egroup
  }{2010}]{mingxing2010}
Mingxing, X., Ming, Y., Jiawen, R., Songtao, A., Jiancheng, K.,  and Dongchen,
  E., 2010, Surface mass balance and ice flow of the glaciers {A}ustre
  {L}ov\'enbreen and {P}edersenbreen, {S}valbard, {A}rctic: Chine Journal of
  Polar Science, {\bf 21}, no. 2, 147--159.

\bibitem[\protect\citeauthoryear{Moore \bgroup et al.\egroup }{1999}]{moore99}
Moore, J.~C., Palli, A., Ludwig, F., Blatter, H., Jania, J., Gadek, B.,
  Glowacki, P., Mochnacki, D.,  and Isaksson, E., 1999, High-resolution
  hydrothermal structure of {H}ansbreen, {S}pitsbergen, mapped by
  ground-penetrating radar: Journal of Glaciology.

\bibitem[\protect\citeauthoryear{Murray and Booth}{2009}]{murray2009}
Murray, T., and Booth, A.~D., 2009, Imaging glacial sediment inclusions in
  3-{D} using ground-penetrating radar at {K}ongsvegen, {S}valbard: Journal of
  Quaternary Science, {\bf 25}, no. 5, 754--761.

\bibitem[\protect\citeauthoryear{Murray \bgroup et al.\egroup
  }{2000}]{murray2000}
Murray, T., Stuart, G.~W., Miller, P.~J., Woodward, J., Smith, A.~M., Porter,
  P.~R.,  and Jiskoot, H., 2000, Glacier surge propagation by thermal evolution
  at the bed: Journal of Geophysical Research.

\bibitem[\protect\citeauthoryear{Murray \bgroup et al.\egroup
  }{2007}]{murray2007}
Murray, T., Booth, A.,  and Rippin, D.~M., 2007, Water-content of glacier-ice:
  Limitations on estimates from velocity analysis of surface ground-penetrating
  radar surveys: Journal of Environmental and Engineering Geophysics, {\bf 12},
  no. 1, 87--99.

\bibitem[\protect\citeauthoryear{Ram\'irez \bgroup et al.\egroup
  }{2001}]{ramirez2001}
Ram\'irez, E., Francou, B., Ribstein, P., Descloitres, M., Guerin, R., Mendoza,
  J., Gallaire, R., Pouyaud, B.,  and Jordan, E., 2001, Small glaciers
  disappearing in the tropical {A}ndes: a case-study in {B}olivia: {G}laciar
  {C}hacaltaya (16$^\circ$ {S}): Journal of Glaciology, {\bf 47}, no. 157,
  187--194.

\bibitem[\protect\citeauthoryear{Rippin \bgroup et al.\egroup
  }{2003}]{rippin2003}
Rippin, D., Willis, I., Arnold, N., Hodson, A., Moore, J., Kohler, J.,  and
  Bj{\"o}rnsson, H., 2003, Changes in geometry and subglacial drainage of
  {M}idre {L}ov{\'e}nbreen, {S}valbard, determined from digital elevation
  models: Earth Surface Processes and Landforms, {\bf 28}, no. 3, 273--298.

\bibitem[\protect\citeauthoryear{Saalmann and Thiedig}{2002}]{saalmann2002}
Saalmann, K., and Thiedig, F., 2002, Thrust tectonics on {B}roggerhalvoya and
  their relationship to the {T}ertiary {W}est {S}pitsbergen {F}old-and-{T}hrust
  {B}elt: Geol. Mag., {\bf 139}, 47--72.

\bibitem[\protect\citeauthoryear{Saintenoy \bgroup et al.\egroup
  }{2011}]{saintenoy2011b}
Saintenoy, A., Friedt, J.-M., Tolle, F., Bernard, E., Laffly, D., Marlin, C.,
  and Griselin, M., June 2011, High density coverage investigation of the
  {A}ustre {L}ov\'enbreen ({S}valbard) using ground-penetrating radar: 6th
  International Workshop on Advanced Ground Penetrating Radar (IWAGPR).

\bibitem[\protect\citeauthoryear{Sandmeier}{2007}]{sandmeier2007}
Sandmeier, K.~J.
\newblock {\bf Reflexw manual, version 4.5}.
\newblock www.sandmeier-geo.de, July 2007.

\bibitem[\protect\citeauthoryear{Stockwell}{1999}]{stockwell1999}
Stockwell, J.~W., May 1999, The {CWP/SU}: {S}eismic {U}n*x {P}ackage: Computers
  and Geosciences, pages 415--419.

\bibitem[\protect\citeauthoryear{Stuart \bgroup et al.\egroup
  }{2003}]{stuart2003}
Stuart, G., Murray, T., Gamble, N., Hayes, K.,  and Hodson, H., 2003,
  {Characterization of englacial channels by ground-penetrating radar: An
  example from {A}ustre Br{\o}ggerbreen, Svalbard}: Journal of Geophysical
  Research, {\bf 108}, no. B11, 2525--+.

\bibitem[\protect\citeauthoryear{Webster and Oliver}{2001}]{webster2001}
Webster, R., and Oliver, M.~A., 2001, Geostatistics for environmental
  scientists ({S}tatistics in {P}ractice): John Wiley and Sons, Brisbane,
  Australia, 1$^{st}$ edition.

\end{thebibliography}
\end{document}